\documentclass[11pt,aps,prd,preprint,superscriptaddress]{revtex4}
\usepackage{revsymb}
\usepackage{amssymb}
\usepackage{amsmath}
\usepackage[dvips]{graphicx}
\newcommand{\be}{\begin{equation}}
\newcommand{\ee}{\end{equation}}
\newcommand{\ben}{\begin{eqnarray}}
\newcommand{\een}{\end{eqnarray}}

\begin{document}
\title{Viscous dark fluid universe}
\date{\today}
\author{W.S. Hip\'{o}lito-Ricaldi\footnote{E-mail: hipolito@ceunes.ufes.br}}
\affiliation{Universidade Federal do Esp\'{\i}rito Santo, Departamento de Ci\^encias Matem\'aticas e Naturais, CEUNES\\
Rodovia BR 101 Norte, km. 60, CEP 29932-540,
S\~ao Mateus, Esp\'{\i}rito Santo, Brazil}
\author{H.E.S.
Velten\footnote{E-mail: velten@cce.ufes.br}}
\affiliation{Universidade Federal do Esp\'{\i}rito Santo,
Departamento
de F\'{\i}sica\\
Av. Fernando Ferrari, 514, Campus de Goiabeiras, CEP 29075-910,
Vit\'oria, Esp\'{\i}rito Santo, Brazil}
\author{W. Zimdahl\footnote{E-mail: winfried.zimdahl@pq.cnpq.br}}
\affiliation{Universidade Federal do Esp\'{\i}rito Santo,
Departamento
de F\'{\i}sica\\
Av. Fernando Ferrari, 514, Campus de Goiabeiras, CEP 29075-910,
Vit\'oria, Esp\'{\i}rito Santo, Brazil}

\begin{abstract}
We investigate the cosmological perturbation dynamics for a universe consisting of pressureless baryonic matter
and a viscous fluid, the latter representing a unified model of the dark sector. In the homogeneous and isotropic background the \textit{total} energy density of this mixture behaves as a generalized Chaplygin gas.
The perturbations of this energy density are intrinsically nonadiabatic and source relative entropy perturbations.
The resulting baryonic matter power spectrum is shown to be compatible with the 2dFGRS and SDSS (DR7) data.
A joint statistical analysis, using also Hubble-function and supernovae Ia data, shows that, different from other studies, there exists a maximum in the probability distribution for a negative present value $q_{0} \approx - 0.53$ of the deceleration parameter.
Moreover, while previous descriptions on the basis of generalized Chaplygin-gas models were incompatible  with  the matter power-spectrum data since they required a much too large amount of pressureless matter, the unified model presented here favors a matter content that is of the order of the baryonic matter abundance suggested by big-bang nucleosynthesis.
\end{abstract}

\maketitle

\section{Introduction}

Since 1998 a huge amount of data has been accumulated which directly or indirectly back up the conclusion, first obtained in \cite{Riess}, that our current Universe entered a phase of accelerated expansion. Direct support is provided by the luminosity-distance data of supernovae of type Ia (SNIa) \cite{SNIa} (but see also \cite{sarkar}), indirect support comes from the anisotropy spectrum of the cosmic microwave background radiation \cite{cmb}, from large-scale-structure data \cite{lss}, from the integrated Sachs--Wolfe effect
\cite{isw}, from baryonic acoustic
oscillations \cite{eisenstein} and from gravitational lensing \cite{weakl}.
Most current cosmological models rely on the assumption that the dynamics of the Universe is described by Einstein's general relativity and a material content that
is dominated by two so far unknown components, pressureless dark matter (DM) and dark energy (DE), a substance equipped with a large negative pressure. For reviews of the actual situation see \cite{rev,pad,dumarev} and references therein.
The preferred model is the $\Lambda$CDM model which also plays the role of a reference model for alternative approaches to the DE problem.
According to the interpretation of the data within this model, our Universe is dynamically dominated by a cosmological constant $\Lambda$ which contributes more than 70\% to the total cosmic energy budget. More than  20\% are contributed by cold dark matter (CDM) and only about 5\% are in the form of conventional, baryonic matter. Because of the cosmological constant problem in its different facets, including the coincidence problem (see, e.g., \cite{straumann,padrev}), a great deal of work was devoted to alternative approaches in which a similar dynamics as that of the $\Lambda$CDM model is reproduced
with a time varying cosmological term, i.e., the cosmological constant is dynamized.
Both DM and DE manifest themselves so far only through their gravitational interaction.
This provides a motivation for approaches in which DM and DE appear as different manifestation of one single
dark-sector component. The Chaplygin-gas model and its different
generalizations  \cite{pasquier,fabris1,bertolami,avelino1,bilic,Finelli1,zimdahl,gorini,NeoN,chaprel} realize this idea. Unified models of the dark sector of this type are attractive since one and the same component behaves as pressureless matter at high redshifts and as a cosmological constant in the long time limit.
While the homogeneous and isotropic background dynamics for the (generalized) Chaplygin gas is well compatible and even slightly favored \cite{colistete} by the data, the study of the perturbation dynamics resulted in problems which apparently ruled out all Chaplygin-gas type models except those that are observationally almost indistinguishable from the $\Lambda$CDM model  \cite{Sandvik}. The point here is that a generally finite adiabatic speed of
sound in generalized Chaplygin gas (GCG) models predicts oscillations (or instabilities) in the power spectrum which are not observed.  Also the analysis of the
anisotropy spectrum of the cosmic microwave background disfavored these models
\cite{Finelli2,Bean}, except possibly for low values of the Hubble
parameter \cite{bento3}.
To circumvent this problem, nonadiabatic perturbations were postulated and designed in a way to make the effective sound speed vanish \cite{ioav,amendola}.
But this amounts to an ad hoc procedure which leaves open the physical origin of nonadiabatic perturbations.
There exists, however a different type of unified model of the dark sector, namely viscous models of the cosmic medium. It was
argued in \cite{antif,NJP}, that a viscous pressure can
play the role of an agent that drives the present
acceleration of the Universe. The option of a
viscosity-dominated late epoch of the Universe with accelerated
expansion was already mentioned in \cite{PadChi}, long before the direct observational evidence through the SN Ia data.
For more recent investigations see, e.g. \cite{rose,Szydlowski,BVM,avelino,barrow,avelino10} and references therein.
In the homogeneous and isotropic background viscous fluids share the same dynamics as GCGs  \cite{rose,Szydlowski,BVM}.
But while perturbations in a (one-component) GCG are always adiabatic, viscous models of the dark sector are intrinsically nonadiabatic.
In a recent paper we showed explicitly that, different from the Chaplygin-gas case, the power spectrum for viscous matter is well behaved and consistent with large-scale-structure data. In particular, it does not suffer from the mentioned oscillation problem \cite{VDF}.
On the other hand, what is observed in the redshift surveys is not the spectrum of the dark-matter
distribution but the baryonic matter spectrum. Including a baryon component into the perturbation dynamics for a universe with a Chaplygin-gas dark sector, it turned out, that the mentioned oscillation within the dark component are not transferred to the baryons \cite{NeoN,chaprel}. The baryonic matter power spectrum is well behaved and consistent with observation. Instead, there appears the new problem that the unified Chaplygin-gas scenario itself is disfavored by the data. It is only if the unified scenario with a fixed pressureless (supposedly) baryonic matter fraction of about $0.043$ (according to the WMAP results) is \textit{imposed} on the dynamics, that consistency with the data is obtained. If the pressureless matter fraction is left free, its best-fit value  is much larger than the baryonic fraction. In fact it becomes even close to unity, leaving only a small percentage for the Chaplygin gas, thus invalidating the entire scenario. In other words, a Chaplygin-gas-based unified model of the dark sector is difficult to reconcile with observations. One may ask now, whether the status of unified models
can again be remedied by replacing the Chaplygin gas by a viscous fluid. It is exactly this question that we are going to investigate in the present paper.
It is our purpose to study cosmological perturbations for a two-component model of baryons and
a viscous fluid, where the latter represents a one-component description of the dark sector.
We shall show that such type of unified model is not only consistent for a fixed fraction of the baryons but also for the case that the matter fraction is left free.  Our analysis demonstrates that the statistically preferred value for the abundance of pressureless matter is compatible with the mentioned baryon fraction $0.043$ that follows from the synthesis of light elements.

The contents of the paper is as follows: in section \ref{The two-component model} we establish our two-component model of a viscous dark component and baryons and discuss its background dynamics.
Section \ref{Perturbations} is devoted to the perturbation dynamics of this mixture. Subsection \ref{total} considers the nonadiabatic total energy-density perturbations, subsection \ref{relative} presents a dynamical equation for the relative entropy perturbations and in subsection \ref{baryons} we obtain the fractional baryonic energy-density perturbations which are shown to be adiabatic at high redshifts.
A numerical integration and tests against data from the matter power spectrum, the Hubble function $H(z)$ and SNIa are given in section \ref{Numerical analysis}, which also
contains a statistical analysis of the validity of the viscous unified model itself.
Finally, section \ref{Discussion} summarizes and discusses our main results.

\section{The two-component model}
\label{The two-component model}

The cosmic medium is assumed  to be describable  by an energy-momentum
tensor
\begin{equation}
T^{ik} = \rho u^{i}u^{k} + p h^{ik} \ ,\qquad h^{ik} = g^{ik} +
u^{i}u^{k}
 \, , \label{Tik}
\end{equation}
which
splits into a matter part $T^{ik}_{M}$ and viscous fluid part $T^{ik}_{V} $,
\begin{equation}
T^{ik} = T^{ik}_{M} + T^{ik}_{V} \ ,\label{T}
\end{equation}
with
\begin{equation}
T^{ik}_{M} = \rho_{M} u^{i}_{M}u^{k}_{M} + p_{M}\left(g^{ik} +
u^{i}_{M}u^{k}_{M}\right) \ ,\qquad T^{ik}_{V} = \rho_{V}
u^{i}_{V}u^{k}_{V} + p_{V}\left(g^{ik} + u^{i}_{V}u^{k}_{V}\right)
\ ,\label{Tmv}
\end{equation}
where the subscript ``M" stands for matter and the subscript ``V" stands for viscous.
The total cosmic fluid is characterized by a four velocity $u^{m}$ while $u^{i}_{M}$ represents
the four velocity of the matter part and $u^{i}_{V}$ represents the four velocity of the viscous fluid.
Energy-momentum conservation is supposed to hold separately for
each of the components,
\begin{equation}
T^{ik}_{M\,;i} = T^{ik}_{V\,;i} = 0 \quad \Rightarrow\quad
T^{ik}_{\ ;i} = 0\ .\label{T;}
\end{equation}
In particular, the energy balances are
\begin{equation}
\rho_{M ,i}u^{i}_{M} + u^{i}_{M ;i}\left(\rho_{M} + p_{M}\right) =
0 \ , \qquad \rho_{V ,i}u^{i}_{V} + u^{i}_{V ;i}\left(\rho_{V} +
p_{V}\right) = 0\ \label{ebalvm}
\end{equation}
and
\begin{equation}
\rho_{,i}u^{i} + u^{i}_{;i}\left(\rho + p\right) = 0 \ ,  \label{ebaltot}
\end{equation}
where (up to first order) $\rho=\rho_M+\rho_V$ and $p=p_M+p_V$.
In general, the four velocities of the components are different. We
shall assume, however, that they coincide in the homogeneous and
isotropic zeroth order,
\begin{equation}
u^{i}_{M} = u^{i}_{V} = u^{i} \qquad \mathrm{(background )}  \ .
\label{}
\end{equation}
Difference will be important only at the perturbative level.

\noindent Let the matter be pressureless and the viscous fluid be characterized by a bulk viscous pressure $p_{V}$,
\begin{equation}
p_{M} = 0\ , \qquad p_{V} = p = - \zeta \Theta\ ,
\label{pmv}
\end{equation}
where $\zeta = $ const and $\Theta = u^{i}_{;i}$ is the fluid
expansion. Under this condition the total pressure coincides with the pressure
of the viscous component.
The total background energy density is $\rho = \rho_{M} + \rho_{V}$,
where
\begin{equation}
\dot{\rho}_{V} + 3 H\left(\rho_{V} + p_{V}\right) = 0 \ , \quad
\dot{\rho}_{M} + 3 H\,\rho_{M}  = 0 \quad\Rightarrow\quad \rho_{M}
= \rho_{M0}a^{-3}
 \, . \label{balB}
\end{equation}
The total energy balance
is $
\dot{\rho} + 3 H\left(\rho + p\right) = 0$.
In the homogeneous and isotropic background one has $\Theta = 3H$,
where $H$ is the Hubble rate. If, moreover, the background is
spatially flat, the Friedmann equation
$3 H^{2} = 8\,\pi\,G\,\rho$
implies $\Theta \propto \rho^{1/2}$, such that
$p = - \zeta \left(24\pi G\right)^{1/2} \rho^{1/2}$. This coincides with the special case $\alpha = - \frac{1}{2}$ for the equation of state $p = -\frac{A}{\rho^{\alpha}}$ of a generalized Chaplygin gas, if we identify
$A = \zeta\,\sqrt{24\,\pi\,G}$.
In terms of the present value $q_{0}$ of the deceleration parameter $q  = - 1 -\frac{\dot{H}}{H^{2}}$ the total energy density can be written as \cite{VDF}
\begin{equation}
\frac{\rho}{\rho_{0}} = \frac{1}{9}\,\left[1 - 2q_{0} + 2 \left(1 +
q_{0}\right)a^{-\frac{3}{2}}\right]^{2} \ ,\quad\Rightarrow\quad
\frac{H}{H_{0}} = \frac{1}{3}\,\left[1 - 2q_{0} + 2 \left(1 +
q_{0}\right)a^{-\frac{3}{2}}\right]\ ,
\label{r/r0q}
\end{equation}
where $\rho_{0}$ and $H_{0}$ denote the present values of  $\rho$ and $H$, respectively.
Since $\rho_{M} =
\rho_{M0}a^{-3}$, we have
$\rho_{V} = \rho -
\rho_{M0}a^{-3}$.
These relations show that it is the total energy density that behaves as a GCG, not the
component $V$. This type of unified model differs from unified models in which the total energy density
is the sum of a GCG and a baryon component. Only if the baryon component is ignored, both
descriptions coincide.
For the total equation of state parameter  we
obtain
\begin{equation}
\frac{p}{\rho} = - \frac{1 - 2q_{0}}{1 - 2q_{0} + 2 \left(1 +
q_{0}\right)a^{-\frac{3}{2}}}
 \ .
\label{p/rq}
\end{equation}
Consequently, in the homogeneous and isotropic background,
a generalized Chaplygin gas with $\alpha = - 1/2$ can be seen as a
unified description of the cosmic medium, consisting of a separately conserved matter component and
a bulk viscous fluid with $\zeta = $ const, where the latter itself represents a unified model of the dark sector.

\section{Perturbations}
\label{Perturbations}
\subsection{Nonadiabatic perturbations of the total density}
\label{total}

The system is characterized by the equations of state (\ref{pmv}). It is expedient to
emphasize that we have neither an equation of state $p_{V} =
p_{V}(\rho_{V})$ nor an equation of state $p = p(\rho)$. It is
only in the spatially flat background when, via Friedmann's
equation, the relation $p = - \zeta \Theta$ reduces to $p \propto
- \rho^{1/2} $ and the corresponding energy density coincides with
the energy density of a GCG.
Neither the component $V$ nor the system as a whole are adiabatic.
Because of $p = - \zeta \Theta$, the pressure perturbation is $\hat{p} = -
\zeta\hat{\Theta}$, where a hat on top of the symbol denotes the (first-order) perturbation of the corresponding quantity.
The nonadiabaticity of the system as a whole
is characterized by
\begin{equation}
\frac{\hat{p}}{\rho + p} - \frac{\dot{p}}{\dot{\rho}}
\frac{\hat{\rho}}{\rho + p} \equiv P - \frac{\dot{p}}{\dot{\rho}} D = 3 H \frac{\dot{p}}{\dot{\rho}}
\left(\frac{\hat{\rho}}{\dot{\rho}} -
\frac{\hat{\Theta}}{\dot{\Theta}}\right)
 \, ,  \label{P-}
\end{equation}
where we have introduced the abbreviations
\begin{equation}
P \equiv \frac{\hat{p}}{\rho + p} \ , \qquad D \equiv
\frac{\hat{\rho}}{\rho + p}\ .\label{PD}
\end{equation}
The quantity (\ref{P-}) is governed by the dynamics of the total energy-density
perturbation $\hat{\rho}$ and by the perturbations $\hat{\Theta}$
of the expansion scalar, which is also a quantity that
characterizes the system as a whole. The behavior of these
quantities is described by the energy-momentum conservation for
the entire system and by the Raychaudhuri equation, respectively. Both of these
equations are coupled to each other.
The remarkable point is that these quantities and, consequently,
the total energy density perturbation, are independent of the
two-component structure of the medium. The reason is the direct relation $\hat{p} = -
\zeta\hat{\Theta}$
between the pressure perturbations and the
perturbations of the expansion scalar.
This is different from  perturbations in a two-component system where each of the components is adiabatic on its own.
It will turn out that the total energy-density perturbations are characterized by a homogeneous
second-order differential equation. These perturbations, which are intrinsically nonadiabatic, then
act as source terms in the evolution equation for
the relative entropy perturbations. The perturbations in the baryon component are obtained
as a combination of the total and the relative entropy perturbations.

The general line element for scalar perturbations is
\begin{equation}
\mbox{d}s^{2} = - \left(1 + 2 \phi\right)\mbox{d}t^2 + 2 a^2
F_{,\alpha }\mbox{d}t\mbox{d}x^{\alpha} +
a^2\left[\left(1-2\psi\right)\delta _{\alpha \beta} + 2E_{,\alpha
\beta} \right] \mbox{d}x^\alpha\mbox{d}x^\beta \ .\label{lineel}
\end{equation}
Since $g_{mn}u^{m}u^{n} = -1$ and also $g_{mn}u_{A}^{m}u_{A}^{n} = -1$, it follows that
\begin{equation}
\hat{u}_0 = \hat{u}^{0} = \hat{u}_{M}^{0} =\hat{u}_{V}^{0}= - \phi \quad \mathrm{and} \quad
a^2\hat{u}^\mu + a^2F_{,\mu} = \hat{u}_\mu \equiv v_{,\mu} \ .
\label{umy}
\end{equation}
The last relation defines the quantity $v$ which will be used to introduce
gauge invariant quantities on comoving ($v=0$) hypersurfaces. Similarly, one defines the corresponding
quantities $v_{M}$ and $v_{V}$ for the components.
These different velocity potentials are related by
\begin{equation}
v_M = v + \frac{\rho_{V} + p_{V}}{\rho + p} \left( v_{M} -
v_{V}\right) \quad \mathrm{and} \quad
v_V = v  - \frac{\rho_{M}}{\rho + p} \left( v_{M} -
v_{V}\right) \ .\label{vv}
\end{equation}
We also introduce the quantity
\begin{equation}
\chi \equiv a^2\left(\dot{E} -F\right) \ .\label{chi}
\end{equation}
The combination $v + \chi$ is gauge invariant.
It is convenient to describe the perturbation dynamics in terms of
gauge invariant quantities which represent perturbations on
comoving hypersurfaces, indicated by a superscript $c$. These are defined as
\begin{equation}
\frac{\hat{\rho}^{c}}{\dot{\rho}} \equiv
\frac{\hat{\rho}}{\dot{\rho}} + v \ , \qquad
\frac{\hat{\Theta}^{c}}{\dot{\Theta}} \equiv
\frac{\hat{\Theta}}{\dot{\Theta}} + v \ , \qquad
\frac{\hat{p}^{c}}{\dot{p}} \equiv \frac{\hat{p}}{\dot{p}} + v \
 \, .  \label{defc}
\end{equation}
For the fractional quantities we introduce the abbreviations
\begin{equation}
D^{c} \equiv \frac{\hat{\rho}^{c}}{\rho + p}\ ,\qquad P^{c} \equiv \frac{\hat{p}^{c}}{\rho + p}\ .\label{Dc}
\end{equation}
In our case we have
\begin{equation}
\frac{\hat{p}}{\dot{p}} = \frac{\hat{\Theta}}{\dot{\Theta}} \qquad
\Rightarrow\qquad \frac{\hat{p}^{c}}{\dot{p}} =
\frac{\hat{\Theta}^{c}}{\dot{\Theta}}
 .  \label{pcThetac}
\end{equation}
In terms of the comoving quantities the total energy and
momentum balances may be combined into (cf. \cite{VDF})
\begin{equation}
\dot{D}^{c} - 3H\,\frac{\dot{p}}{\dot{\rho}} \, D^{c}  +
\hat{\Theta}^c =0 \ . \label{dotD}
\end{equation}
The expansion scalar $\Theta $ is governed by the Raychaudhuri
equation,
\begin{equation}
\dot{\Theta} + \frac{1}{3}\Theta^{2} + 2\left(\sigma^{2} -
\omega^{2}\right) - \dot{u}^{a}_{;a} + 4\pi\,
G\,\left(\rho + 3 p\right) = 0 \ . \label{ray}
\end{equation}
Up to first order the perturbed Raychaudhuri equation can be
written in the form
\begin{equation}
\dot{\hat{\Theta}}^c + 2H\hat{\Theta}^c +
\frac{1}{a^{2}}\Delta P^{c} + \frac{3\gamma}{2}H^{2} \, D^{c}
= 0\ . \label{pertRay}
\end{equation}
It is through the
Raychaudhuri equation that the pressure gradient comes into play:
\begin{equation}
P^{c} = \frac{p}{\gamma \rho}\,\frac{\hat{\Theta}^{c}}{\Theta}\ ,
\quad \Rightarrow\quad
P^{c}  =
\frac{1}{2\gamma}\frac{p^{2}}{\rho^{2}} D^{c} - \frac{p}{3\gamma
\rho H}\dot{D}^{c}\ ,\label{PcDc}
\end{equation}
where $\gamma = 1 + \frac{p}{\rho}$.
The pressure perturbation consists of a term which is proportional
to the total energy-density perturbations $D^{c}$ (notice that the
factor in front of $D^{c}$ is positive), but additionally of a
term proportional to the time derivative $\dot{D}^{c}$ of $D^{c}$.
The relation between pressure perturbations $P^{c}$ and
energy perturbations $D^{c}$ is no longer simply algebraic, equivalent to a
(given) sound-speed parameter as a factor relating the two. The
relation between them becomes part of the dynamics. In a sense,
$P^{c}$ is no longer a ``local" function of $D^{c}$ but it
is a function of the derivative $\dot{D}^{c}$ as well \cite{essay}. This is equivalent to
$\hat{p} = \hat{p}(\hat{\rho}, \dot{\hat{\rho}})$. It is only for
the background pressure that the familiar dependence $p = p(\rho)$
is retained.
As already mentioned, the two-component structure of the medium is not relevant here.

Introducing now
\begin{equation}
\delta \equiv \gamma D^{c} =\frac{\hat{\rho}^{c}}{\rho} \,
\ ,  \label{deltac}
\end{equation}
and changing from the variable  $t$ to $a$,
Eqs.~(\ref{dotD}) and (\ref{pertRay}) may be combined to yield the second-order equation
\begin{equation}
\delta'' + f\left(a\right)\delta' + g\left(a\right) \,\delta = 0 \ ,\label{dddshort}
\end{equation}
where $\delta' \equiv \frac{d \delta}{d a}$ and the coefficients $f$ and $g$ are
\begin{equation}
f\left(a\right) = \frac{1}{a}\,\left[\frac{3}{2} - 6\frac{p}{\rho}  - \frac{1}{3}\frac{p}{\gamma\rho}\,\frac{k^{2}}{H^{2}
a^{2}}\right]  \label{f}
\end{equation}
and
\begin{equation}
g\left(a\right) = - \frac{1}{a^{2}}\,\left[\frac{3}{2}  + \frac{15}{2}
\frac{p}{\rho} - \frac{9}{2}\,\frac{p^{2}}{\rho^{2}}
- \frac{1}{\gamma}\frac{p^{2}}{\rho^{2}} \frac{k^{2}}{H^{2}
a^{2}}\right]\ , \label{g}
\end{equation}
respectively.
Equation (\ref{dddshort}) coincides with the corresponding equation for the one-component case in \cite{VDF}.

\subsection{Relative entropy perturbations}
\label{relative}

Alternatively to relation (\ref{P-}),  the deviation from adiabaticity in a
two-component system with components $M$ and $V$  is

\begin{eqnarray}
\frac{\hat{p}}{\rho + p} -
\frac{\dot{p}}{\dot{\rho}}\frac{\hat{\rho}}{\rho + p}  = P^{c} - \frac{\dot{p}}{\dot{\rho}}D^{c} &=&
 \frac{\rho_{V} + p_{V}}{\rho + p}
\left(\frac{\hat{p}_{V}}{\rho_{V} + p_{V}} -
\frac{\dot{p}_V}{\dot{\rho}_V}\frac{\hat{\rho}_{V}}{\rho_{V} +
p_{V}}  \right)\nonumber\\
\ \nonumber\\&& + \frac{\rho_{M} \left(\rho_{V} + p_{V}\right)} {\left(\rho +
p\right)^2} \frac{\dot{p}_V}{\dot{\rho}_V}
\left[\frac{\hat{\rho}_{V}}{\rho_{V} + p_{V}} -
\frac{\hat{\rho}_{M}}{\rho_{M}} \right] \ .\label{}
\end{eqnarray}
\noindent Solving this for the nonadiabatic part of component $V$
yields
\begin{equation}
\frac{\hat{p}_{V}}{\rho_{V} + p_{V}} -
\frac{\dot{p}_{V}}{\dot{\rho}_{V}}\frac{\hat{\rho}_{V}}{\rho_{V} +
p_{V}} = \frac{\rho + p}{\rho_{V} + p_{V}}
\left[P^{c} - \frac{\dot{p}}{\dot{\rho}}D^{c} - 3 H
\frac{\dot{p}}{\dot{\rho}}\frac{\dot{\rho}_{M}}{\dot{\rho}}
\left(\frac{\hat{\rho}_{M}}{\dot{\rho}_{M}} -
\frac{\hat{\rho}_{V}}{\dot{\rho}_{V}}\right) \right] \
.\label{hatp2}
\end{equation}
The perturbed energy balances for the components ($A = M, V$) are
\begin{equation}
\left(\frac{\hat{\rho}_A}{\rho_A + p_A}
\right)^{\displaystyle\cdot} + 3H \left(\frac{\hat{p}_A}{\rho_A +
p_A} - \frac{\dot{p}_A}{\dot{\rho}_A}\frac{\hat{\rho}_A}{\rho_A +
p_A} \right) -3\dot{\psi} + \frac{1}{a^2}\left (\Delta v_{A}
+\Delta \chi\right) =0 \ .\label{}
\end{equation}
Obviously, the combination (\ref{hatp2}) enters the energy balance of the viscous component.
Subtracting the balance of fluid $M$ from the balance of fluid $V$ and
using (\ref{hatp2}) it follows that
\begin{eqnarray}
\left(\frac{\hat{\rho}_{V}}{\rho_{V} + p_{V}} -
\frac{\hat{\rho}_{M}}{\rho_{M}}\right)^{\displaystyle
\cdot} &+& 3 H\left\{\frac{\rho + p}{\rho_{V} + p_{V}} \left[P^{c} - \frac{\dot{p}}{\dot{\rho}}D^{c}
- 3 H
\frac{\dot{p}}{\dot{\rho}}\frac{\dot{\rho}_{M}}{\dot{\rho}}
\left(\frac{\hat{\rho}_{M}}{\dot{\rho}_{M}} -
\frac{\hat{\rho}_{V}}{\dot{\rho}_{V}}\right)
\right]\right\}\nonumber\\
&& \qquad\qquad\qquad\qquad\qquad \qquad\qquad + \frac{1}{a^{2}}
\Delta \left(v_{V} - v_{M}\right) = 0
 \ .\label{dotdiff}
\end{eqnarray}
To deal with the term that contains the difference $v_{V} - v_{M}$ of the velocity potentials of the components, we implement
the momentum balances which imply ($A = M, V$)
\begin{equation}
\frac{\hat{p}_A}{\rho_A + p_A} + \frac{\dot{p}_A}{\rho_A + p_A}v_A
+ \dot{v}_A + \phi=0\ . \label{}
\end{equation}
With $p_{M} = 0$, the definition for $P^{c}$
in (\ref{Dc}) and with  (\ref{vv})
we arrive at
\begin{equation}
\left(v_{V} - v_{M}\right)^{\displaystyle\cdot} = - \frac{\rho +
p}{\rho_{V} + p_{V}} P^{c} - 3H
\frac{\dot{p}}{\dot{\rho}}\frac{\rho_{M}}{\rho_{V} +
p_{V}}\left(v_{M} - v_{V}\right)
 \ .\label{vv-vm}
\end{equation}
Introducing relative entropy perturbations by the usual definition
\begin{equation}
S_{MV} \equiv \frac{\hat{\rho}_{M}}{\rho_{M}} -
\frac{\hat{\rho}_{V}}{\rho_{V} + p_{V}}  \ ,\label{SVM}
\end{equation}
differentiating equation (\ref{dotdiff}) and combining the result with equation (\ref{vv-vm}) and with (\ref{dotdiff}) again,
we obtain the inhomogeneous second-order equation
\begin{equation}
S_{VM}'' + r(a)S_{VM}' + s(a) S_{VM} = c(a) \delta' + d(a)\delta
 \
\label{ppS}
\end{equation}
with the coefficients
\begin{equation}
r(a) = \frac{1}{a}\left[\frac{3}{2} - \frac{3}{2}\frac{p}{\rho}  -
3\frac{p}{\rho}\frac{\rho_{M}}{\rho_{V} +
p}\right]
 \ ,
\label{r(a)}
\end{equation}
\begin{equation}
s(a) = - \frac{3}{a^{2}}
\frac{p}{\rho}\frac{\rho_{M}}{\rho_{V} + p}\left[1 +
\frac{3}{4}\frac{p}{\rho}\right]
 \ ,
\label{s(a)}
\end{equation}
\begin{equation}
c(a) = \frac{1}{a}\left[\frac{3}{\gamma}\frac{p}{\rho_{V} +
p}\,\left(1 +
\frac{p}{2 \rho} + \left(1 + \frac{p}{\gamma
\rho}\right)\frac{k^{2}}{9 H^{2}a^{2}}\right)\right]
 \
\label{c(a)}
\end{equation}
and
\begin{equation}
d(a) = \frac{9}{2\gamma a^{2}}\frac{p}{\rho_{V} +
p}\,\left[\left(1
-\frac{p}{\rho}\right)\left(1 + \frac{p}{2 \rho}\right) -
2\frac{p}{\rho}\left(1 + \frac{p}{\gamma
\rho}\right)\frac{k^{2}}{9
H^{2}a^{2}}\right]
 \ .
\label{d(a)}
\end{equation}
The set of equations (\ref{ppS}) and (\ref{dddshort}) contains the
entire perturbation dynamics of the system. At first, the homogeneous Eq.~(\ref{dddshort}) for $\delta$ has to be solved. Subsequently, once $\delta$ is known, Eq.~(\ref{ppS}) determines the relative entropy perturbations.

\subsection{Baryon density perturbations}
\label{baryons}

The quantity relevant for the observations is the
fractional perturbation $\delta_{M} \equiv \frac{\hat{\rho}_{M}^{c}}{\rho_{M}}$ of the energy density of the baryons.  This
quantity  is obtained from the total fractional density $\delta$, determined through (\ref{dddshort}), and
the relative entropy perturbations $S_{VM}$, determined through (\ref{ppS}),
by
\begin{equation}
\delta_{M} = \frac{1}{\gamma}\left[\delta - \frac{\rho_{V} +
p}{\rho}S_{VM}\right] \ ,
 \label{del1}
\end{equation}
with
\begin{equation}
\frac{\rho_{V} + p}{\rho} = \frac{2 \left(1 +
q_{0}\right)a^{-3/2} \left[1 - 2q_{0} +
2 \left(1 + q_{0}\right)a^{-3/2}\right]
-9\Omega_{M0}a^{-3}}{\left[1 - 2q_{0} + 2
\left(1 +
q_{0}\right)a^{-3/2}\right]^{2}}
 \ ,
\label{r2+p/r}
\end{equation}
where we have introduced the present value of the matter fraction
$\Omega_{M0} \equiv \frac{8\pi G}{3H_{0}^{2}}\rho_{M0}$.
Assuming $H_{0}$ to be given, the free parameters of the system are $q_{0}$ and $\Omega_{M}$.

At early times, i.e. for small scale factors $a \ll 1$, the equation (\ref{dddshort}) has the asymptotic form
\begin{eqnarray}\label{EarlyViscous}
\qquad \qquad \delta'' + \frac{3}{2a} \,\delta' - \frac{3}{2a^2}\,\delta =0 \,,\qquad \qquad  (a \ll 1)
\end{eqnarray}
independent of $q_0$ and for all scales.
The solutions of (\ref{EarlyViscous}) are
\begin{eqnarray} \label{asympsol}
\delta(a\ll 1)= c_1 a + c_2 a^{-3/2} \,,
\end{eqnarray}
where $c_1$ and $c_2$ are integration constants.
The nonadiabatic contributions to the total density perturbations are negligible at high redshifts \cite{VDF}.

For $a \ll 1$ the coefficients $s(a)$, $c(a)$ and $d(a)$ in (\ref{ppS}) become negligible and
$r(a)\rightarrow \frac{3}{2}$. Eq.~(\ref{ppS}) then reduces to
\begin{equation}
S_{VM}'' + \frac{3}{2a}S_{VM}'  = 0\,,\qquad \qquad  (a \ll 1)
 \ .
\label{ppS1}
\end{equation}
It has the solution $S_{VM} = $ const $=0$. From the definition (\ref{SVM}) we find that at high redshifts
\begin{equation}
S_{MV} = \frac{\hat{\rho}_{M}}{\rho_{M}} -
\frac{\hat{\rho}_{V}}{\rho_{V}}\,,\qquad \qquad  (a \ll 1)  \ ,\label{SVM1}
\end{equation}
since $\frac{p}{\rho_{V}} \ll 1$ under this condition.
Consequently, there are neither nonadiabatic contributions to the total energy-density fluctuations nor
relative entropy perturbations and we have purely adiabatic perturbations $\delta_{M} = \delta$ at $a \ll 1$.
This allows us to relate our model to the $\Lambda$CDM model at early times.
We shall use the fact that the matter power spectrum for the
$\Lambda$CDM model is well fitted by the BBKS transfer
function \cite{bbks}.
Integrating the $\Lambda$CDM model back from today to a distant past, say $z = 1.000$, we
obtain the shape of the transfer function at that moment.
The
spectrum determined in this way is then used as initial
condition for our viscous model. This procedure is similar to that described in
more detail in references \cite{sola,saulo}.

\section{Statistical analysis}
\label{Numerical analysis}

To estimate the free parameters of our model we perform a Bayesian analysis and construct the corresponding
probability distribution functions. At first we consider large-scale-structure data from the 2dFGRS \cite{cole} and
SDSS DR7 \cite{sdss} programs.
The matter power spectrum is defined by
\begin{equation}
P_k=\left|\delta_{M,k}\right|^{2}\ ,
\end{equation}
where $\delta_{M,k}$ is the Fourier component of the density contrast $\delta_{M}$.
Generally, for a set of free parameters $\left\{{\bf p}\right\}$, the agreement between the theoretical prediction and observations is assessed by minimizing the quantity
\begin{equation}
\chi^{2}\left({\bf p}\right)=\frac{1}{Nf}\sum_{i}\frac{\left[P^{th}_{i}({\bf p}) - P^{obs}_{i}({\bf p})\right]^{2}}{\sigma_{i}^{2}},
\end{equation}
where Nf means the number of degrees of freedom in the analysis. The quantities $P^{th}_{i}$ and $P^{obs}_{i}$ are the theoretical and the observed values, respectively, of the power spectrum and $\sigma_{i}$ denotes the error for the data point $i$. With the help of $\chi^2$ we then construct the probability density function (PDF)
\begin{equation}
P = \mathcal{B} \,e^{-\frac{\chi^{2}(\bf{p})}{2}}\ ,
\end{equation}
where $\mathcal{B}$ is a normalization constant.

\begin{center}
\begin{figure}[!h]
\hspace{0cm}
\begin{minipage}[t]{0.45\linewidth}
\includegraphics[width=\linewidth]{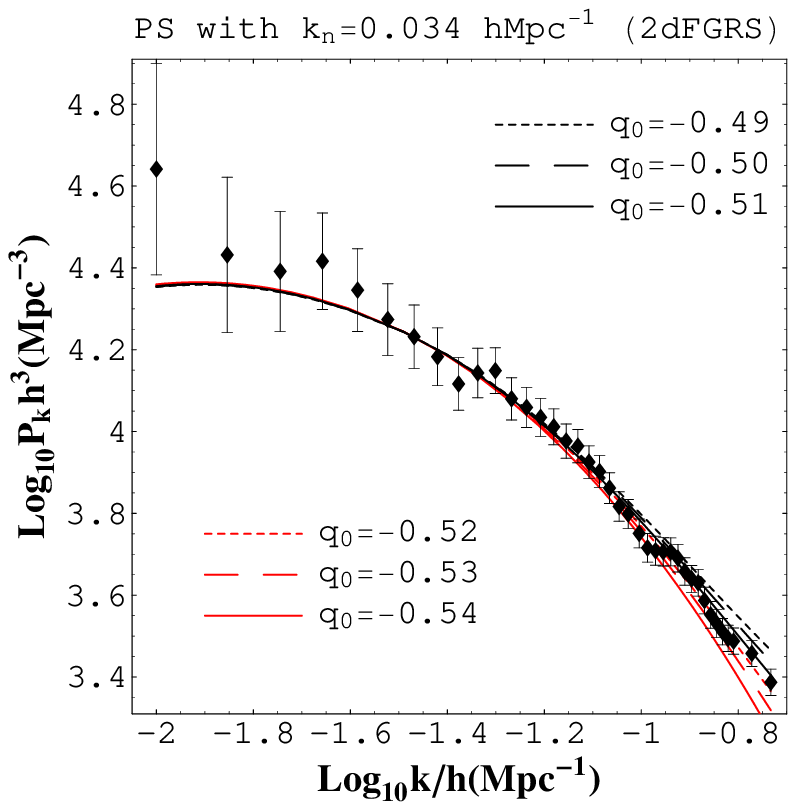}
\end{minipage} \hfill
\begin{minipage}[t]{0.45\linewidth}
\includegraphics[width=\linewidth]{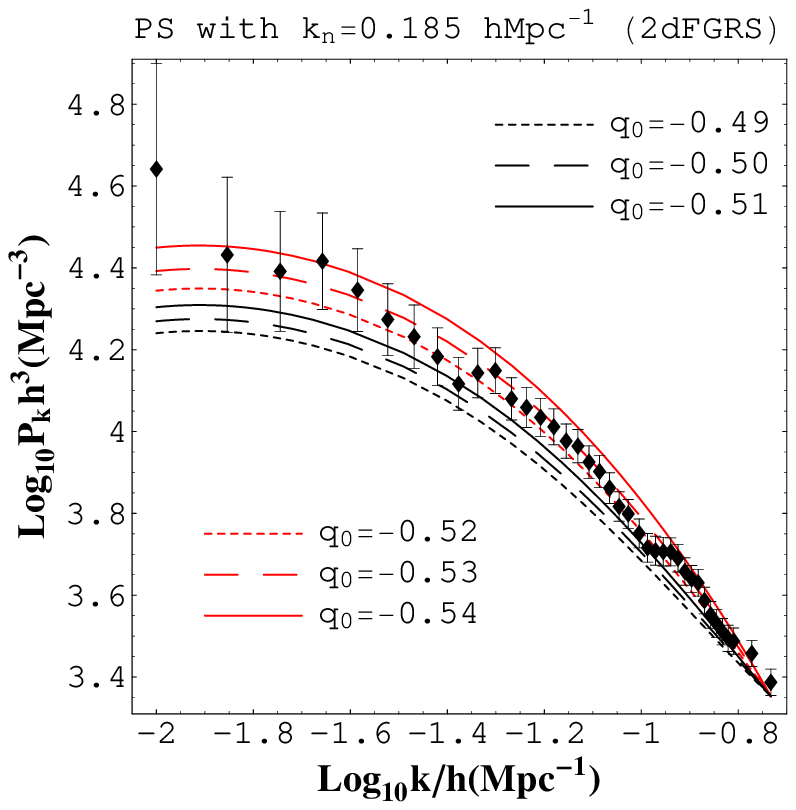}
\end{minipage} \hfill
\begin{minipage}[t]{0.45\linewidth}
\includegraphics[width=\linewidth]{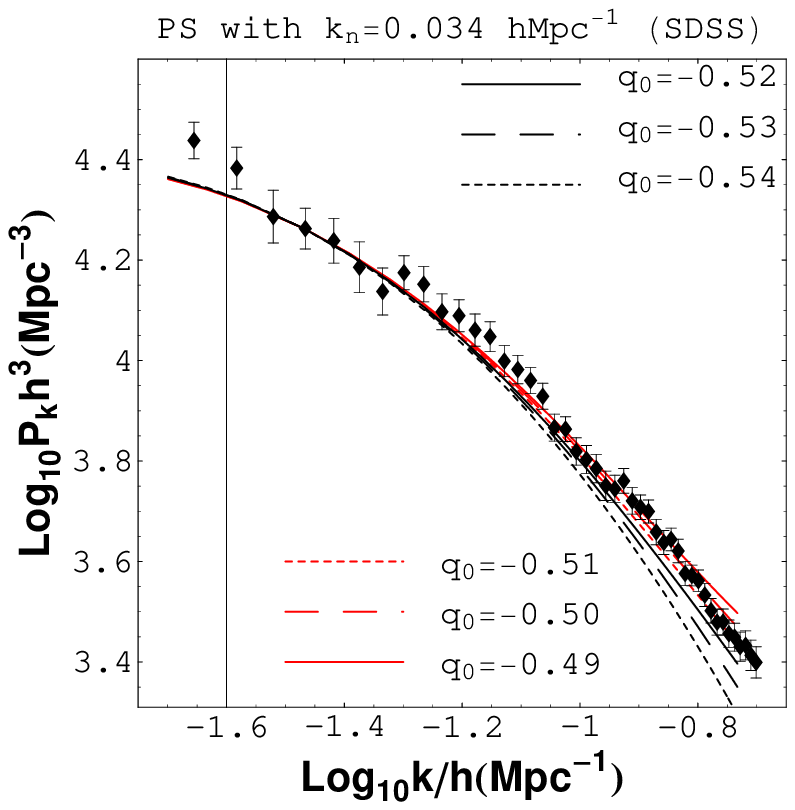}
\end{minipage} \hfill
\begin{minipage}[t]{0.45\linewidth}
\includegraphics[width=\linewidth]{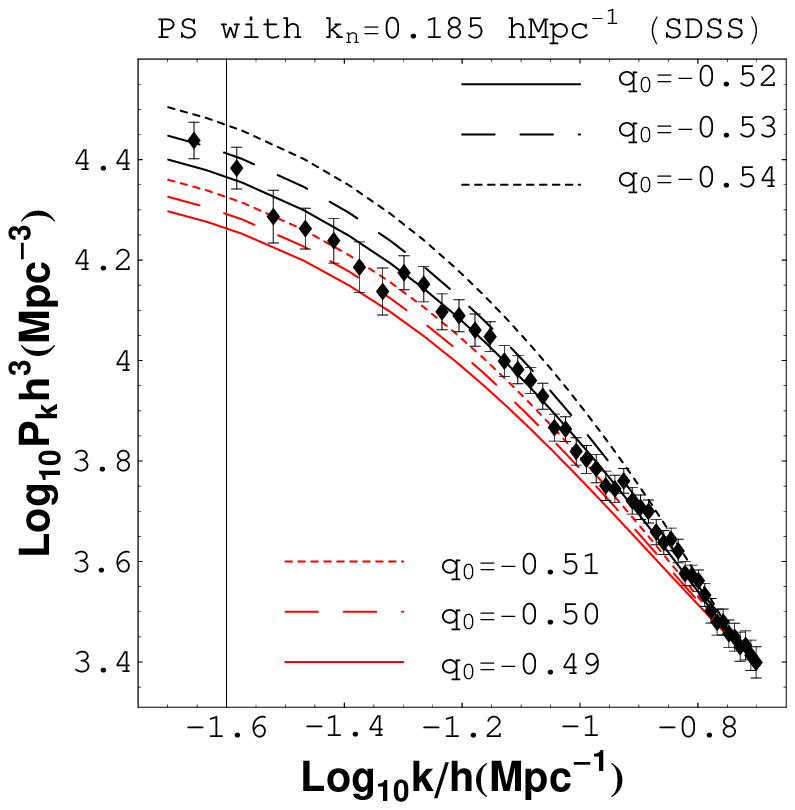}
\end{minipage} \hfill
\caption{{\protect\footnotesize Power spectra (PS) normalized at $k_{n}=0.034 hMpc^{-1}$ (left panels) and at $0.185 hMpc^{-1}$ (right panels) for different negative values of $q_{0}$. The top panels compare the PS with the 2dFGRS data, the bottom panels with the SDSS DR7 data.}}
\label{2}
\end{figure}
\end{center}

\begin{center}
\begin{figure}[!h]
\hspace{0cm}
\begin{minipage}[t]{0.45\linewidth}
\includegraphics[width=\linewidth]{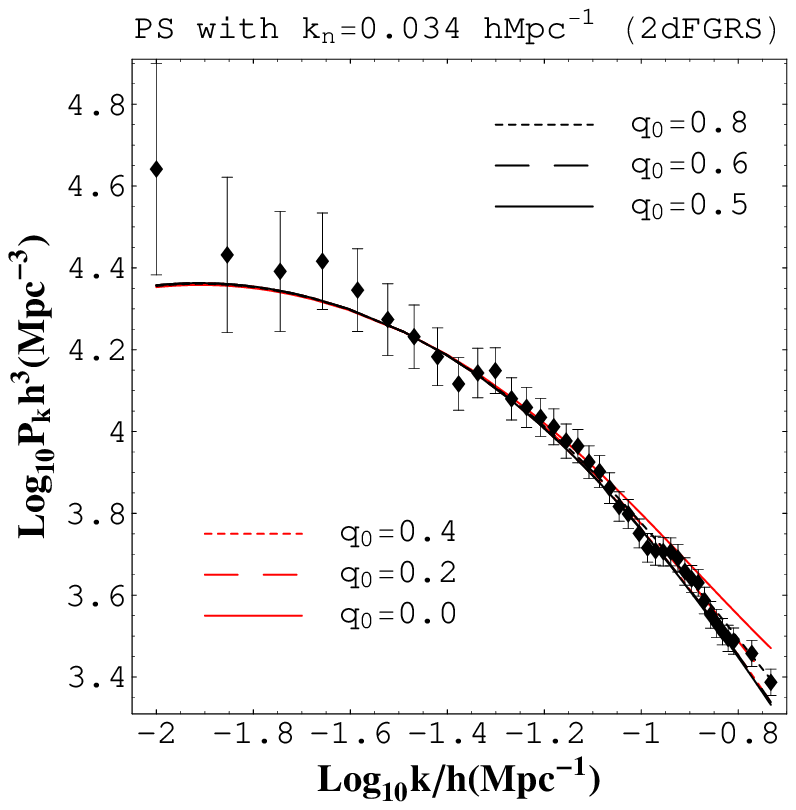}
\end{minipage} \hfill
\begin{minipage}[t]{0.45\linewidth}
\includegraphics[width=\linewidth]{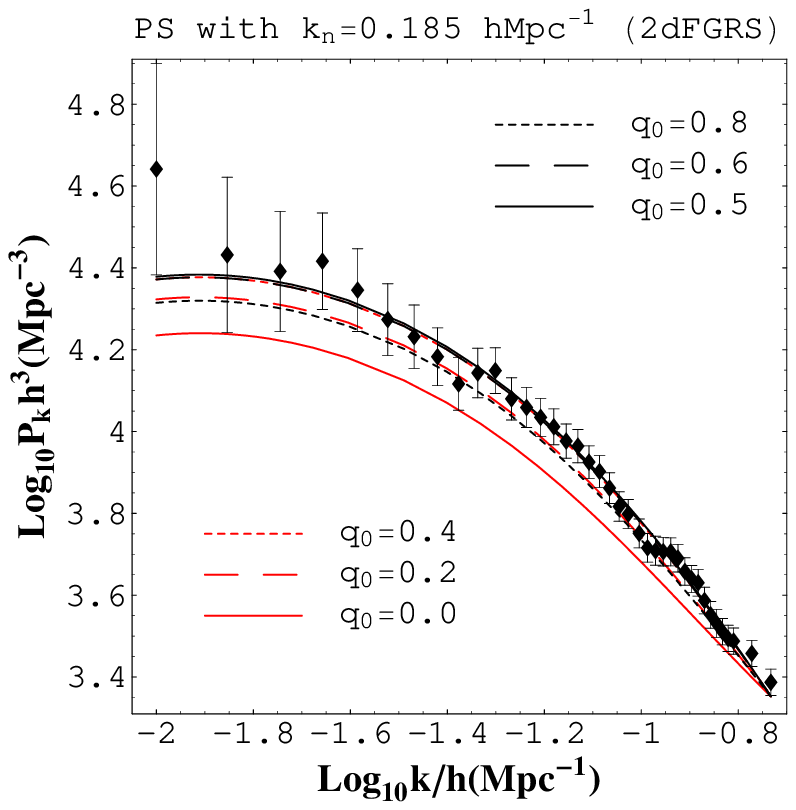}
\end{minipage} \hfill
\begin{minipage}[t]{0.45\linewidth}
\includegraphics[width=\linewidth]{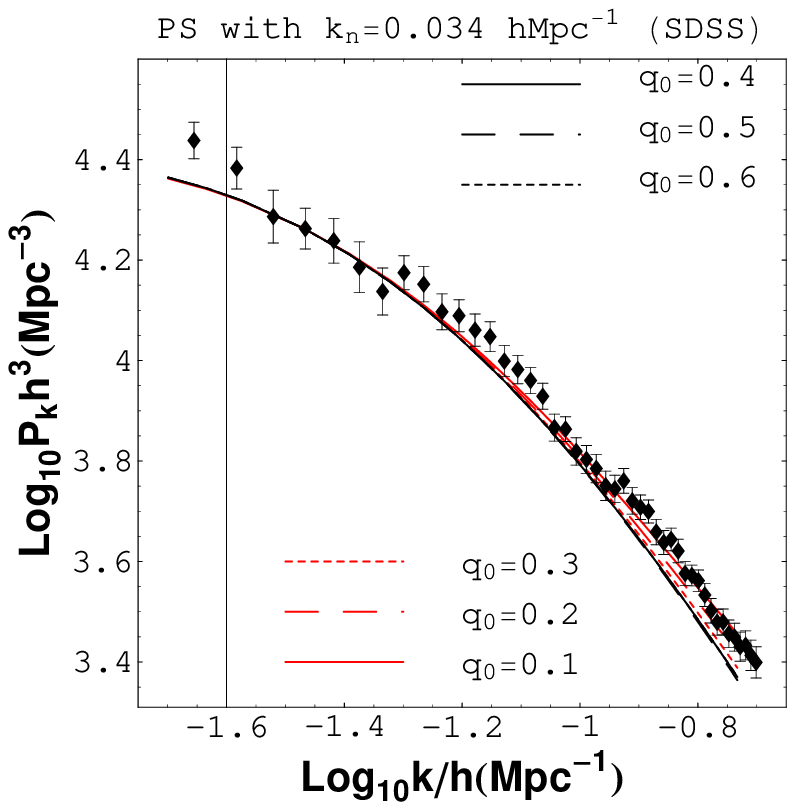}
\end{minipage} \hfill
\begin{minipage}[t]{0.45\linewidth}
\includegraphics[width=\linewidth]{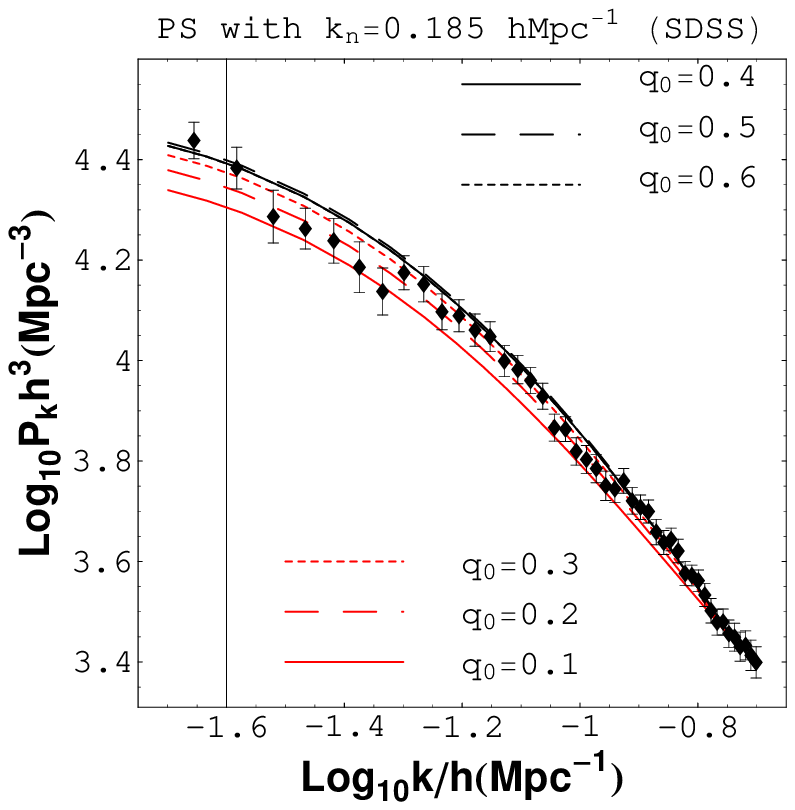}
\end{minipage} \hfill
\caption{{\protect\footnotesize Power spectra (PS) normalized at $k_{n}=0.034 hMpc^{-1}$ (left panels) and at $0.185 hMpc^{-1}$ (right panels) for different positive values of $q_{0}$. The top panels compare the PS with the 2dFGRS data, the bottom panels with the SDSS DR7 data.}}
\label{3}
\end{figure}
\end{center}

To test our model against the observed power-spectra data we consider the following two situations.
(i) We assume the matter component to be entirely baryonic with a fraction $\Omega_{M0} = 0.043$ as suggested by the WMAP data. Fixing also $H_{0} = 72$, a value favored by these data as well, the only remaining free parameter is $q_{0}$. This will provide us with information about the preferred value(s) of $q_{0}$ for the unified dark-sector model.
(ii) We leave the matter fraction free, thus admitting that the matter component is not only made up by the baryons. This is equivalent to allow for a separate DM component in addition to the contribution effectively  accounted for by the viscous fluid.
This additional freedom is used to test our unified model of the dark sector itself. The unified model can be regarded as favored by the data if the PDF for the matter fraction is large around the value that characterizes the baryon fraction. If, on the other hand, the PDF is largest at a substantially higher value, the unified model has to be regarded as disfavored.
The PDF for case (i) is shown in figure \ref{1}. We obtain two regions with high probability for $q_{0}$,
one of them with a pronounced peak around $q_{0} \approx -0.53$, implying accelerated expansion.
The other one, which is of the same hight, has $q_{0} > 0$ and is compatible with an Einstein-de Sitter universe. The appearance of a maximum of the PDF in the region $q_{0}< 0$ is neither observed in Chaplygin-gas scenarios
nor in our previously studied one-component viscous model \cite{VDF}. The difference to the latter might appear surprising since $q_{0}$ characterizes the system as a whole and the addition of a small fraction of baryons should, at the first glance, not have a large impact on the total dynamics. However, it is not the background dynamics that counts here. In the present case the PDF for $q_{0}$ is inferred from power-spectra data that are related to the fluctuations $\delta_{M}$, while in the one-component model these data were related to the fluctuations $\delta$ of the total energy density. As relation~(\ref{del1}) shows,  $\delta_{M}$ and $\delta$ may be very different in general. The appearance of a maximum for $q_{0}< 0$ means, that the results of a first-order analysis may well be compatible with the results for the background, which imply $q_{0}< 0$.
We consider this an advantage over Chaplygin-gas models, for which there has always been a tension between the results in the background
and those on the perturbative level \cite{NeoN,chaprel}.
Figure \ref{2} (Figure \ref{3}) shows the theoretically obtained spectrum for various negative (positive) values of $q_{0}$ together with the power-spectrum data points.
To better illustrate the relation between the predictions of the model and the observations, two different normalization wave numbers, $k_{n}=0.034hMpc^{-1}$ and $k_{n}=0.185hMpc^{-1}$, have been chosen, but our statistical results do not depend on a specific normalization.

\begin{center}
\begin{figure}[!h]
\hspace{0cm}
\begin{minipage}[t]{0.4\linewidth}
\includegraphics[width=\linewidth]{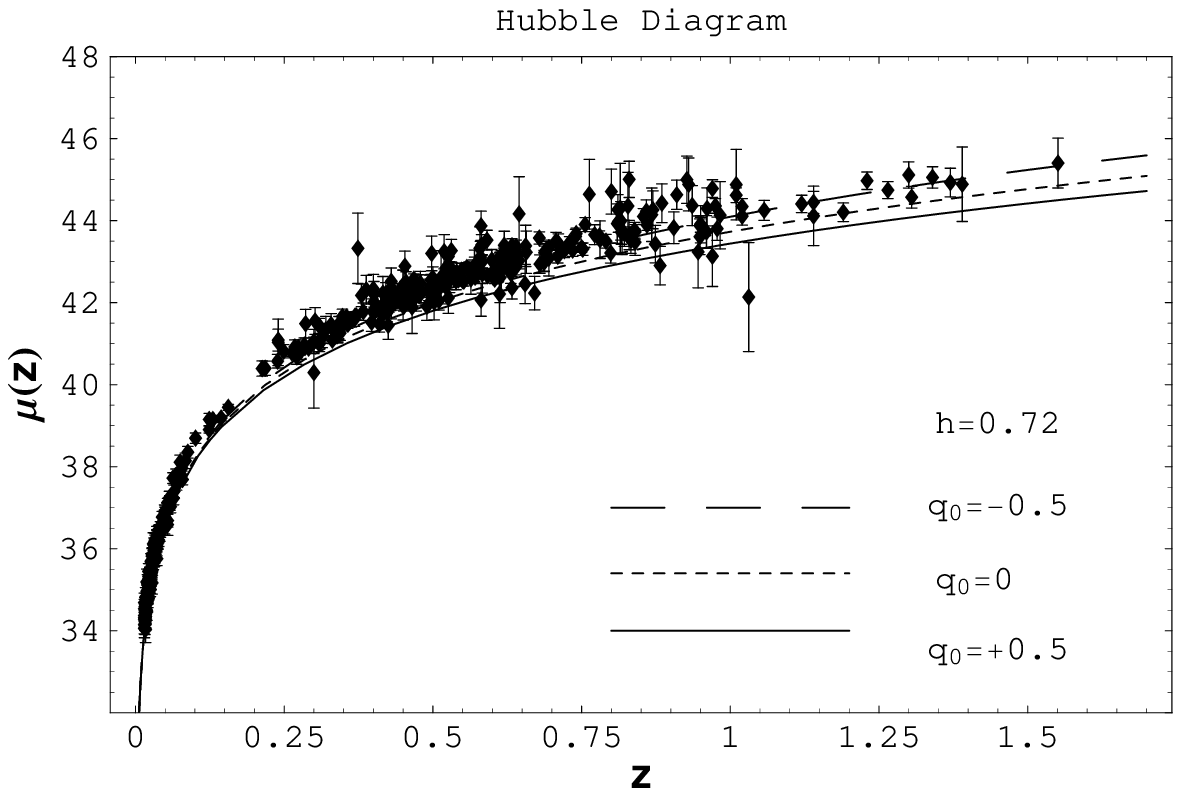}
\end{minipage} \hfill
\begin{minipage}[t]{0.25\linewidth}
\includegraphics[width=\linewidth]{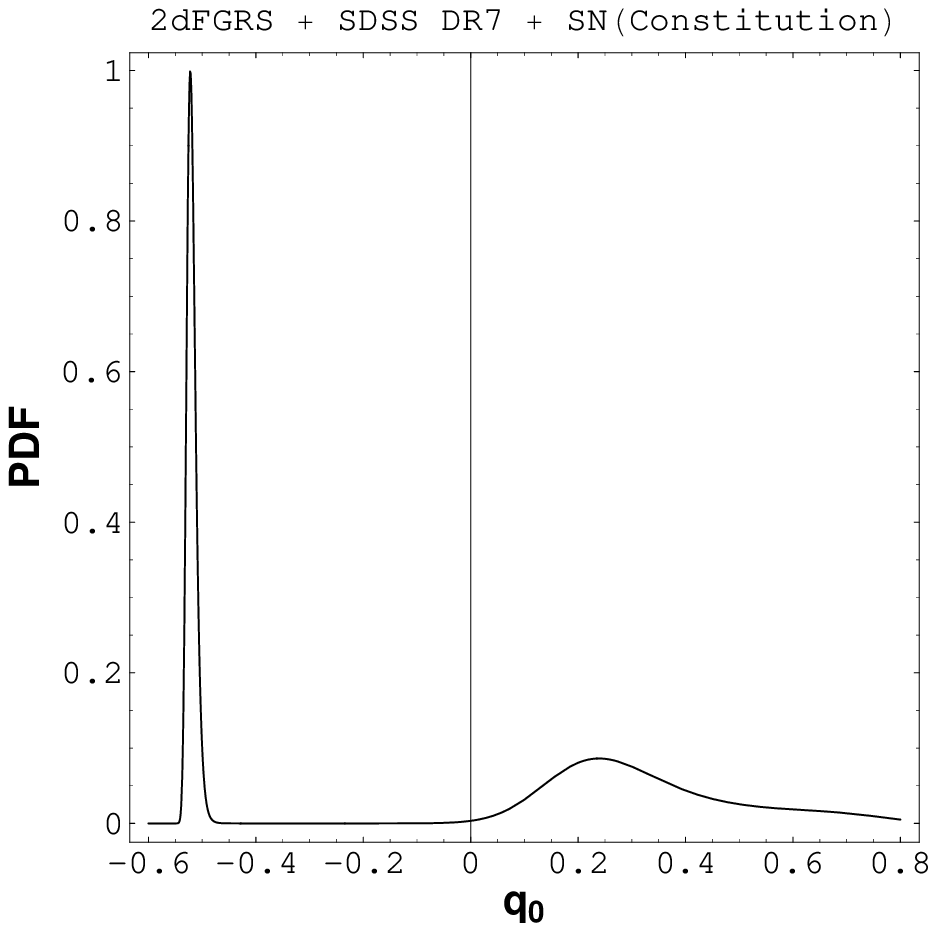}
\end{minipage} \hfill
\begin{minipage}[t]{0.25\linewidth}
\includegraphics[width=\linewidth]{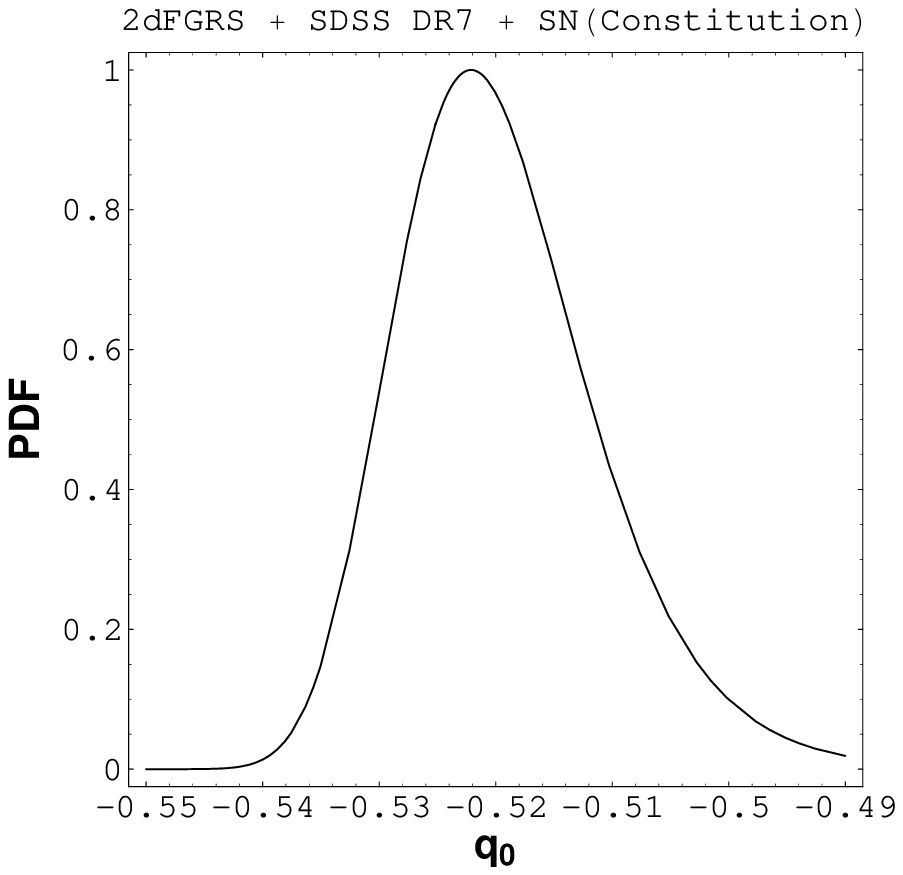}
\end{minipage} \hfill
\caption{{\protect\footnotesize Left panel: Hubble diagram, center panel: PDF for $q_{0}$, based on a joint analysis of PS and SN data. The right panel magnifies the maximum for $q_{0}<0$.}}
\label{SN}
\end{figure}
\end{center}

\begin{center}
\begin{figure}[!h]
\hspace{0cm}
\begin{minipage}[t]{0.4\linewidth}
\includegraphics[width=\linewidth]{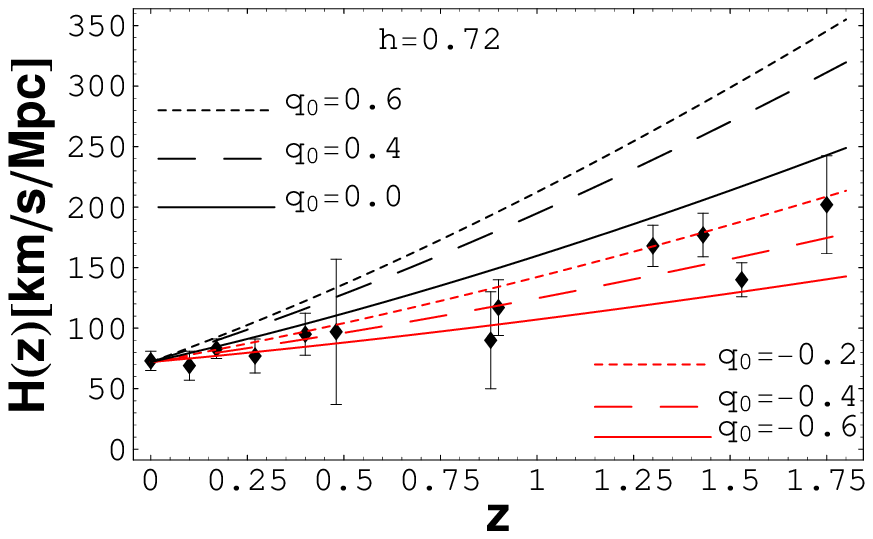}
\end{minipage} \hfill
\begin{minipage}[t]{0.3\linewidth}
\includegraphics[width=\linewidth]{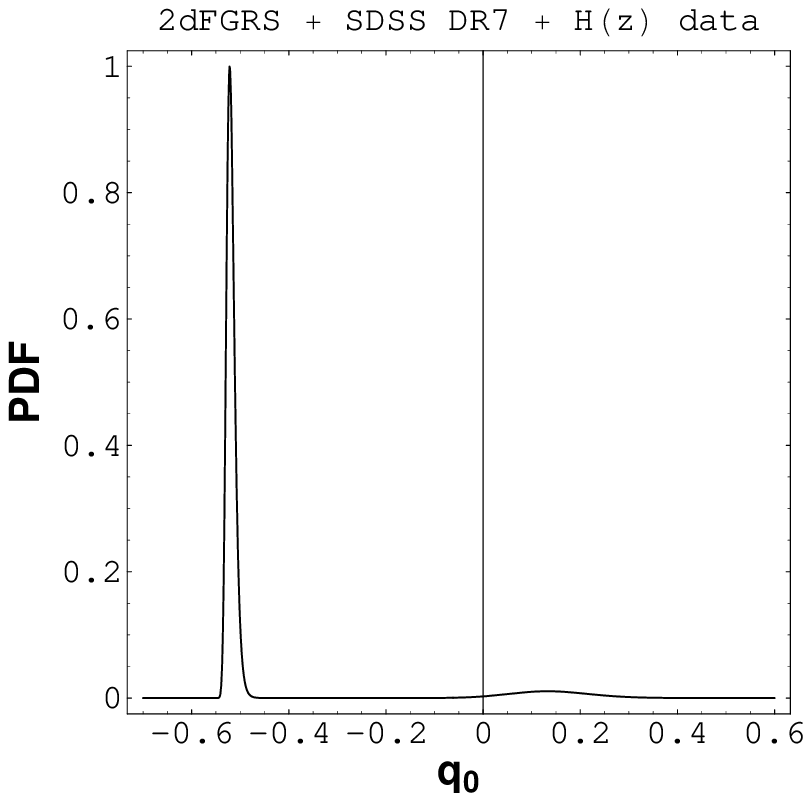}
\end{minipage} \hfill
\caption{{\protect\footnotesize Left panel: Hubble parameter as a function of the redshift for different values of $q_{0}$. Right panel: one-dimensional PDF for $q_{0}$, based on a joint analysis of PS and H(z) data.}}
\label{H}
\end{figure}
\end{center}

In order to break the degeneracy between the high-probability regions in Fig.~\ref{1} we include
information from the Constitution set of SNIa (see the last reference in \cite{SNIa}) and from the recent H(z) data from \cite{stern}.
The results from the joint analysis with SNIa data are shown in Fig.~\ref{SN}, while  Fig.~\ref{H} depicts the corresponding
PDF for $q_{0}$, based on the joint analysis with the H(z) data.
In Fig.~\ref{SN} the total $\chi^{2}$ is calculated from
$\chi^{2} = \chi^{2}_{2dFGRS} + \chi^{2}_{SDSS} + \chi^{2}_{SNIa}$, in Fig.~\ref{H} from $\chi^{2} = \chi^{2}_{2dFGRS} + \chi^{2}_{SDSS} + \chi^{2}_{H}$.

\begin{center}
\begin{figure}[!h]
\hspace{0cm}
\begin{minipage}[t]{0.3\linewidth}
\includegraphics[width=\linewidth]{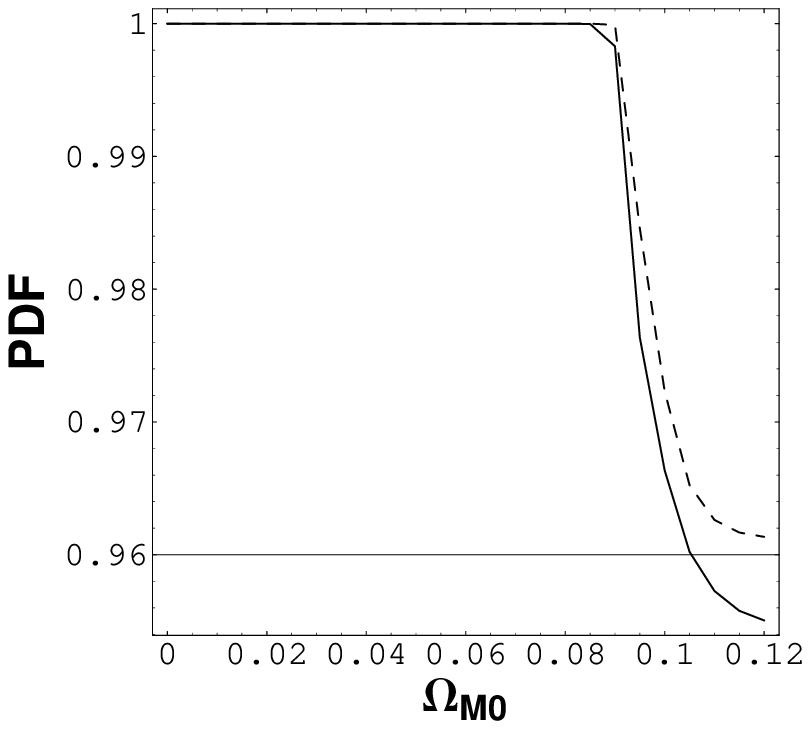}
\end{minipage} \hfill
\begin{minipage}[t]{0.3\linewidth}
\includegraphics[width=\linewidth]{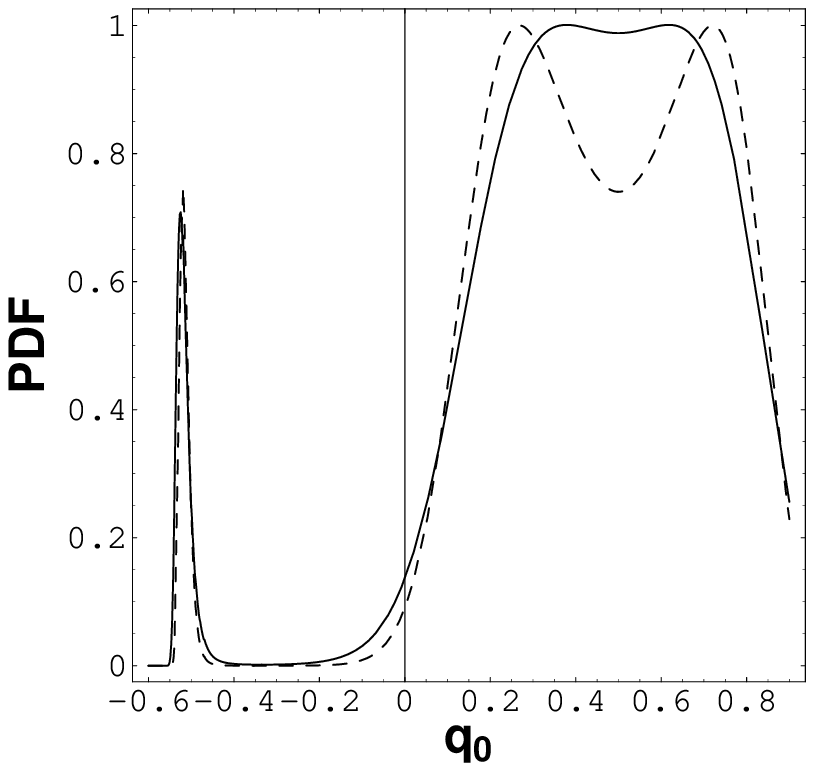}
\end{minipage} \hfill
\begin{minipage}[t]{0.3\linewidth}
\includegraphics[width=\linewidth]{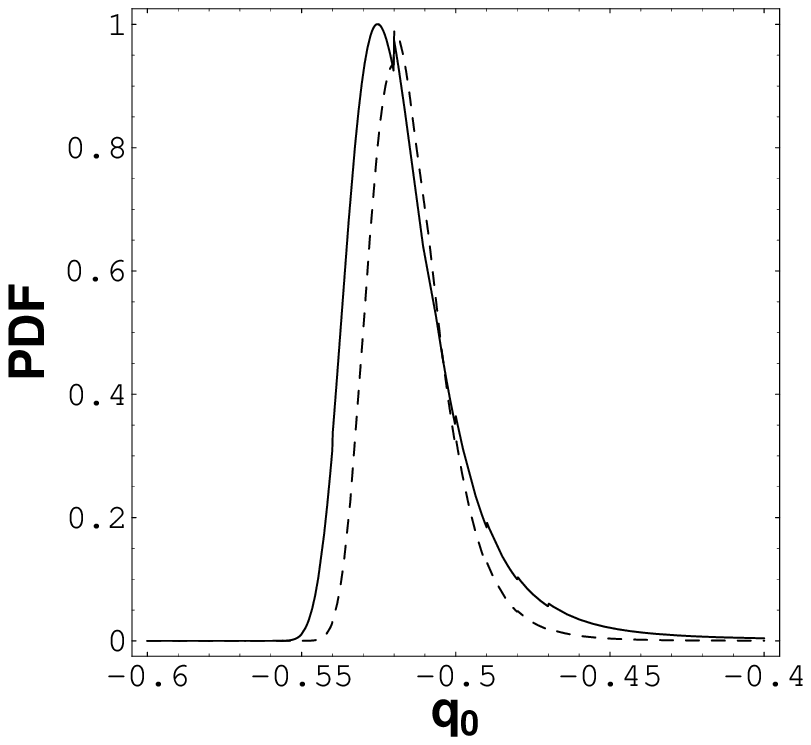}
\end{minipage} \hfill
\caption{{\protect\footnotesize PDF for the pressureless component $\Omega_{M0}$ (left) and for the deceleration parameter $q_{0}$ (center), using the 2dFGRS data (solid lines) and the SDSS DR7 data (dashed lines). The right picture is a normalized amplification of the peak for $q_{0}<0$ in the central panel.}}
\label{2pm}
\end{figure}
\end{center}

For case (ii) we have both $q_{0}$ and $\Omega_{M0}$ as free parameters. The results of the statistical analysis are shown in figure \ref{2pm}. Most importantly, there is a high probability for small values of the matter fraction $\Omega_{M0}$, including the WMAP value $\Omega_{M0} = 0.043$. According to our previously mentioned criteria this means, the unified viscous model
is indeed preferred by the data. This is in striking contrast to unified Chaplygin-gas models which have high probabilities close to $\Omega_{M0} = 1$, thus apparently invalidating the idea of a unified description
of dark matter and dark energy \cite{NeoN,chaprel}.

\begin{center}
\hspace{0cm}
\begin{figure}[!h]
\hspace{0cm}
\begin{minipage}[t]{0.4\linewidth}
\includegraphics[width=\linewidth]{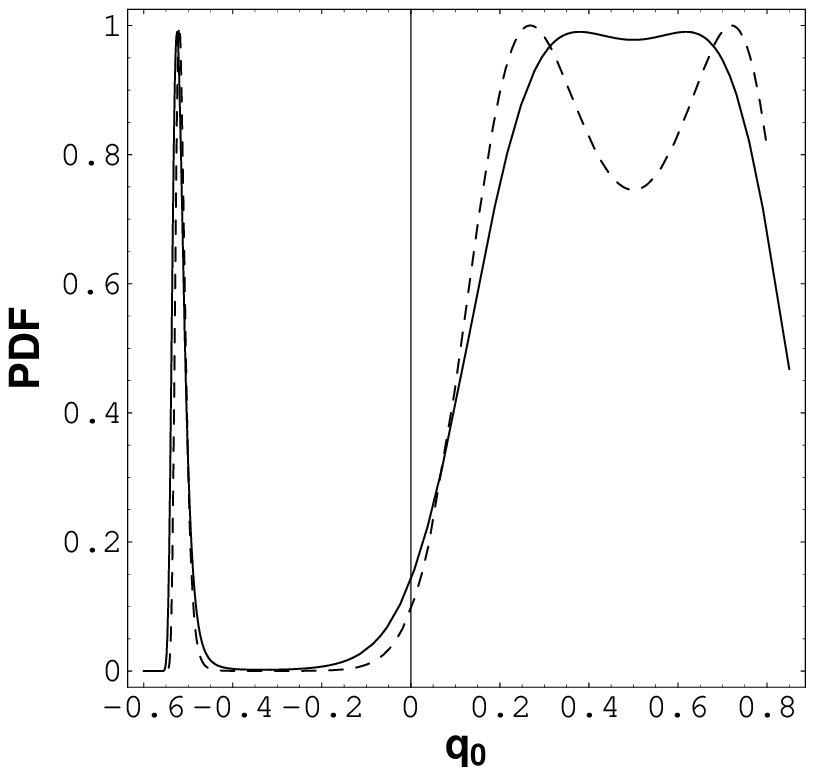}
\end{minipage} \hfill
\begin{minipage}[t]{0.4\linewidth}
\includegraphics[width=\linewidth]{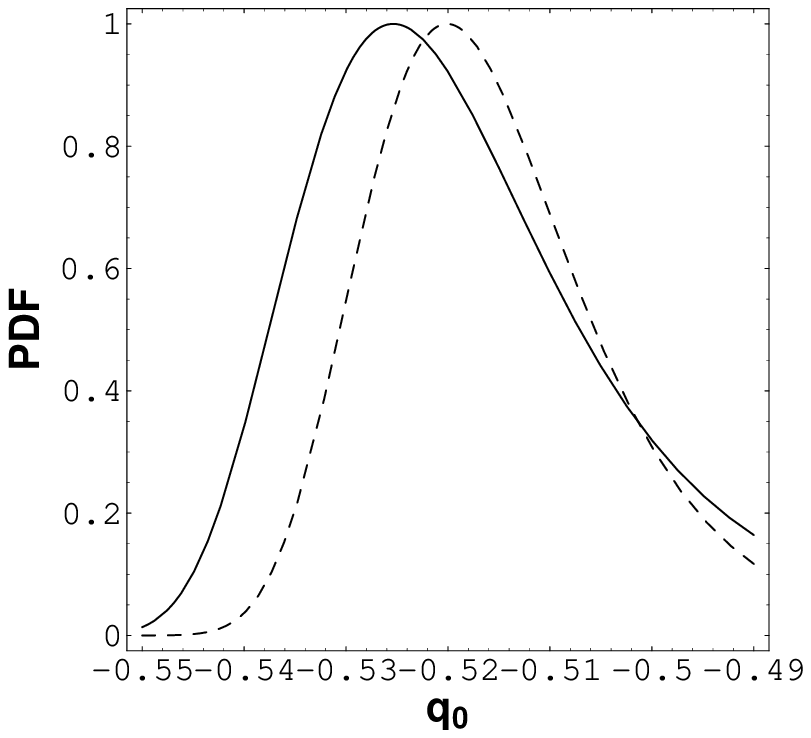}
\end{minipage} \hfill
\caption{{\protect\footnotesize One-dimensional PDF for $q_{0}$ resulting from the 2dFGRS data (solid curve) and from the SDSS DR7 data (dashed curve). The right picture is an amplification of the peak in the region $q_{0}<0$}.}
\label{1}
\end{figure}
\end{center}

\section{Discussion and Conclusions}
\label{Discussion}

It is well known that, in a homogeneous, isotropic and spatially flat background universe, there exists an equivalence between viscous and generalized Chaplygin-gas models for a unified description of the cosmological dark sector \cite{Szydlowski,BVM}. The cosmic substratum at the present time is then approximated as a mixture of one of these dark components and baryons.
The novel approach presented here is based on the fact that also the two-component system of a bulk viscous fluid and a separately conserved baryon component behaves in the background as a generalized Chaplygin gas with
$\alpha = -\frac{1}{2}$. The total energy-density perturbations, however, are intrinsically nonadiabatic and coincide with those of a one-component viscous fluid, investigated in earlier work \cite{VDF}. While the baryon component may be considered dynamically negligible in the background, the situation is different on the perturbative level, since the observed matter agglomerations are related to baryonic density fluctuations.
These fluctuations are obtained from a combination of the said nonadiabatic total energy density perturbations and relative entropy perturbations in the two-component system where the former source the latter.
The observed matter-power spectrum  is well reproduced. There do not appear oscillations or instabilities which have plagued adiabatic Chaplygin-gas models \cite{Sandvik}.
Our present results improve the findings of a previous one-component analysis in which no baryons were included \cite{VDF}.
At first, the probability distribution for the deceleration parameter has a maximum at $q_{0} \approx -0.53$ which partially removes the degeneracy of previous studies which, taken at face value, were incompatible with an accelerated expansion and thus in obvious tension with results for the background.
Perhaps still more important is the test of the unified model itself. Many investigations on approaches with a unified dark sector fix the pressureless matter component to be that of the favored (by the WMAP data) baryon fraction and then check whether or not the resulting dynamics can reproduce the observations. This corresponds to the strategy (i) in the previous section. But this is only part of the story since it does not say
anything on how probable the division of the total cosmic substratum into roughly 96\% of a dark substance and roughly 4\% of pressureless matter is. To decide this question, one has to consider the pressureless matter fraction as a free parameter and to find out which abundance is actually favored by the data. Our analysis in point (ii) was devoted to this task and revealed that the matter fraction probability is indeed highest for values smaller than roughly 8\%. This is a result in favor of the unified viscous model. We recall that a corresponding analysis for a Chaplygin gas results in values close to unity \cite{chaprel} which seems to rule out such type of approaches.
The present viscous model, on the other hand, remains an option for a unified description of the dark sector, at least as far as the matter power spectrum is concerned.\\

\ \\
\noindent
{\bf Acknowledgement:}  We thank FAPES and CNPq (Brazil) for
financial support. WSHR is grateful to FAPESB and to S. Carneiro
for interesting discussions and for their hospitality during his stay at
the Institute of Physics at UFBA.



\begin{thebibliography}{99}
\bibitem{Riess} A.G. Riess {\em et al.}, Astron. J. \textbf{116}, 1009 (1998); S.
Perlmutter {\em et al.}, Astrophys. J. \textbf{517}, 565 (1999).
\bibitem{SNIa}
J.L. Tonry {\em et al.}, Astrophys. J. \textbf{594}, 1 (2003);
M.V. John, Astrophys. J. \textbf{614}, 1 (2004); P. Astier {\em et
al.}, J. Astron. Astrophys. \textbf{447}, 31 (2006); A.G. Riess
{\em et al.}, astro-ph/0611572; D. Rubin {\em et al.}, arXiv:0807.1108; M. Hicken {\em et al.}, Astrophys.J. \textbf{700}, 1097 (2009), arXiv:0901.4804.
\bibitem{sarkar} S. Sarkar, Gen. Relativ. Gravit. \textbf{40}, 269 (2008).
\bibitem{cmb}
S. Hanany {\em et al.}, Astrophys. J. Lett. \textbf{545}, L5 2000;
C.B. Netterfield {\em et al.}, Astrophys. J. \textbf{571}, 604
(2002); E. Komatsu {\em et al.}, Astrophys. J. Suppl. \textbf{180}, 330 (2009), arXiv:0803.0547.
\bibitem{lss}
M. Colless {\em et al.}, Mon. Not. R. Astron. Soc. \textbf{328},
1039 (2001); M. Tegmark {\em et al.}, Phys. Rev. D \textbf{69},
103501 (2004); S. Cole {\em et al.}, Mon. Not. R. Astron. Soc.
\textbf{362}, 505 (2005); V. Springel, C.S. Frenk, and S.M.D.
White, Nature (London) \textbf{440}, 1137 (2006).
\bibitem{isw}
S. Boughn and R. Chrittenden, Nature (London) \textbf{427}, 45
(2004); P. Vielva, E. Mart\'{\i}nez--Gonz\'{a}lez, and M. Tucci,
Mon. Not. R. Astron. Soc. \textbf{365}, 891 (2006).
\bibitem{eisenstein} D.J. Eisenstein {\em et al.},  Ap.J. \textbf{633}, 560 (2005), arXiv:astro-ph/0501171.
\bibitem{weakl}
C.R. Contaldi, H. Hoekstra, and A. Lewis, Phys. Rev. Lett.
\textbf{90}, 221303 (2003).
\bibitem{rev}
E.J. Copeland, M. Sami and S. Tsujikawa, Int.J.Mod.Phys.D15, 1753
(2006).
\bibitem{pad} T. Padmanabhan, Gen. Relativ. Gravit. \textbf{40}, 529 (2008).
\bibitem{dumarev} R. Durrer and R. Maartens, Gen. Relativ. Gravit. \textbf{40}, 301 (2008).
\bibitem{straumann} N. Straumann, astro-ph/0203330.
\bibitem{padrev} T. Padmanabhan, Phys.Rept. \textbf{380}, 235 (2003); hep-th/0212290.
\bibitem{pasquier}  A. Yu. Kamenshchik, U. Moschella and V. Pasquier,
Phys.Lett. \textbf{B511}, 265 (2001).
\bibitem{fabris1}  J.C. Fabris, S.V.B. Gon\c{c}alves e P.E. de Souza, Gen.
Rel. Grav. \textbf{34}, 53 (2002).
\bibitem{bertolami}  M. C. Bento, O. Bertolami and A. A. Sen, Phys. Rev.
\textbf{D66}, 043507 (2002).
\bibitem{avelino1} L.M.G. Be\c{c}a, P.P. Avelino, J.P.M. de
Carvalho and C.J.A.P. Martins, Phys. Rev. {\bf 67}, 101301(R) (2003).
\bibitem{bilic} N. Bilic, G. P. Tupper, and R. D. Viollier, Phys. Lett. \textbf{B535},
17 (2002); N. Bilic, R. J. Lindebaum, G. P. Tupper, and
R. D. Viollier, J. Cosmol. Astropart. Phys. \textbf{11} (2004) 008;
M. C. Bento, O. Bertolami, and A. A. Sen, Phys. Rev. D
\textbf{70}, 083519 (2004); R. R. R. Reis, M. Makler, and I. Waga,
Phys. Rev. D \textbf{69}, 101301 (2004); R. A. Sussman,
arXiv:0801.3324; N. Bilic, G. P. Tupper, and R. D.
Viollier, arXiv: 0809.0375.
\bibitem{Finelli1} D. Carturan and F. Finelli,
Phys. Rev. D 68, 103501 (2003).
\bibitem{zimdahl}  W. Zimdahl and J.C. Fabris, Class. Quant. Grav. \textbf{22}, 4311 (2005).
\bibitem{gorini} V. Gorini, A.Y. Kamenshchik, U. Moschella, O. F.
Piatella, and A. A. Starobinsky, J. Cosmol. Astropart.
Phys. \textbf{02}. 016 (2008).
\bibitem{NeoN} J.C. Fabris, S.V.B. Gon\c{c}alves, H.E.S. Velten and W. Zimdahl,
Phys. Rev. D \textbf{78}, 103523 (2008).
\bibitem{chaprel} J.C. Fabris, H.E.S. Velten and W. Zimdahl,
Phys. Rev. D \textbf{81}, 087303 (2010).
\bibitem{colistete}  R. Colistete Jr, J. C. Fabris, S.V.B. Gon\c{c}alves and
P.E. de Souza, Int. J. Mod. Phys. \textbf{D13}, 669 (2004); R.
Colistete Jr., J. C. Fabris and S.V.B. Gon\c{c}alves, Int. J. Mod.
Phys. \textbf{D14},
775 (2005); R. Colistete Jr. and J. C. Fabris, Class. Quant. Grav. \textbf{22}%
, 2813 (2005); R. Colistete Jr. and R. Giostri, \textit{BETOCS
using the 157 gold SNe Ia Data : Hubble is not humble},
arXiv:astro-ph/0610916.
\bibitem{Sandvik} H.B. Sandvik, M. Tegmark, M. Zaldariaga and I. Waga,
Phys. Rev. D 69, 123524 (2004).
\bibitem{Finelli2} L. Amendola, F. Finelli, C. Burigana and D. Carturan,
JCAP 0307, 005 (2003).
\bibitem{Bean}
R. Bean and O. Dor\'e, Phys. Rev. D 68, 023515  (2003).
\bibitem{bento3} M.C. Bento, O. Bertolami and A.A. Sen,
Phys. Lett. B 575, 172 (2003).
\bibitem{ioav}  R.R.R. Reis, I. Waga, M.O. Calv\~ao e S.E. Jor\`as, Phys.
Rev. \textbf{D68}, 061302 (2003).
\bibitem{amendola} L. Amendola, I. Waga and F. Finelli,
JCAP \textbf{0511}, 009 (2005).
\bibitem{antif}
W. Zimdahl, D.J. Schwarz,  A.B. Balakin, and D. Pav\'{o}n, Phys.
Rev. D \textbf{64}, 063501 (2001).
\bibitem{NJP} A.B. Balakin, D. Pav\'{o}n, D.J. Schwarz, and  W.
Zimdahl, NJP \textbf{5}, 85.1 (2003).
\bibitem{PadChi}  T. Padmanabhan and S. M. Chitre, Phys. Lett. A \textbf{120}, 433 (1987).
\bibitem{rose}  J.C. Fabris, S.V.B. Gon\c{c}alves and R. de S\'a Ribeiro,
Gen. Rel. Grav. \textbf{38}, 495 (2006).
\bibitem{Szydlowski}  M. Szyd{\l}owski and O. Hrycyna,
Ann.Phys. \textbf{322}, 2745 (2007).
\bibitem{BVM} R. Colistete Jr., J.C. Fabris, J. Tossa and W.
Zimdahl, Phys. Rev. {\bf D76}, 103516 (2007).
\bibitem{avelino} A. Avelino and U. Nucamendi, JCAP \textbf{0904} (2009) 006.
\bibitem{barrow} B. Li and J.D. Barrow, Phys. Rev. \textbf{D79}, 103521 (2009).
\bibitem{avelino10} A. Avelino and U. Nucamendi, JCAP \textbf{1008} (2010) 006.
\bibitem{VDF} W.S. Hip\'{o}lito-Ricaldi, H.E.S. Velten  and W. Zimdahl,
JCAP \textbf{06} (2009) 016.
\bibitem{essay} W. Zimdahl, Int. J. Mod. Phys. D (IJMPD) \textbf{17},
651 (2008) (arXiv:0705.2131).
\bibitem{bbks} J.M. Bardeen, J.R. Bond, N. Kaiser and A.S. Szalay,
Astrophys. J. {\bf 304}, 15 (1986); J. Martin, A. Riazuelo and M.
Sakellariadou, Phys. Rev. {\bf D61}, 083518 (2000).
\bibitem{sola} J.C. Fabris, I.L. Shapiro and J. Sol\`a, JCAP {\bf
0702}, (2007) 016.
\bibitem{saulo} H.A. Borges, S. Carneiro, J.C. Fabris and C.
Pigozzo, Phys. Rev. {\bf D77}, 043513 (2008).
\bibitem{cole} S. Cole et al., Mon. Not. R. Astron. Soc. \textbf{362}, 505 (2005).
\bibitem{sdss} B.A. Reid et al., Mon. Not. R. Astron. Soc. \textbf{404}, 60 (2010).
\bibitem{stern} D. Stern, R. Jimenez, L. Verde, M. Kamionkowski and S. A. Stanford,
JCAP {\bf 1002}, 008 (2010).



\end{thebibliography}
\end{document}